\documentclass[12pt]{article}
\usepackage{graphicx}
\title{Resolving the black-hole information paradox \\ 
by treating time on an equal footing with space}
\author{Hrvoje Nikoli\'c \\
Theoretical Physics Division, Rudjer Bo\v{s}kovi\'{c} Institute, \\
P.O.B. 180, HR-10002 Zagreb, Croatia \\
{\normalsize hrvoje@thphys.irb.hr} \\
\makebox[1in]{} \\
}
\date{\today}
\begin{document}
\maketitle
\begin{abstract}
Pure states in quantum field theory can be represented by many-fingered block-time wave functions, 
which treat time on an equal footing with space and make the notions of ``time evolution''
and ``state at a given time'' fundamentally irrelevant. Instead of information destruction 
resulting from an attempt to use a ``state at a given time'' to describe semi-classical  
black-hole evaporation, the full many-fingered block-time wave function of the universe
conserves information by describing the correlations of outgoing Hawking particles in the future with ingoing Hawking particles in the past.
\end{abstract}
\vspace*{0.5cm}

\section{Introduction}

The semi-classical description of black-hole evaporation \cite{hawk1}
predicts that the final state after the complete
evaporation cannot be represented by a pure state \cite{hawk2}.
A transition from a pure to a non-pure (i.e., mixed) state contradicts
unitarity of quantum mechanics and leads also to other pathologies \cite{banks}.
Many approaches to restore a pure-state description at late times have been 
attempted, but none of them seems to be
completely satisfying (for reviews see, e.g., \cite{revije}).

\begin{figure}[t]
\includegraphics[width=4cm]{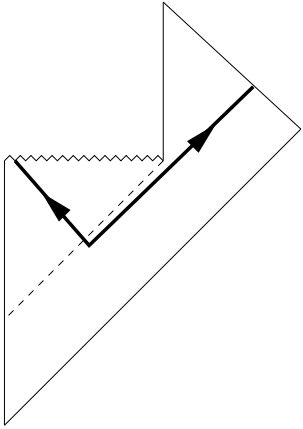}
\caption{\label{fig1}
Penrose diagram of a completely evaporating black hole. The lines with arrows represent
a Hawking pair of particles.}
\end{figure}

To overcome this problem, we start with the observation that all these previous approaches
(with a notable exception in \cite{hartle}) 
share one common assumption: that the quantum state (either pure or mixed) should
be a function of time, or more generally, a functional of the spacelike hypersurface.
Indeed, such an assumption is deeply rooted in our intuitive understanding of the concept
of time, according to which universe evolves with time. Yet, such a view of time
does not seem to be compatible with the 
classical theory of relativity (both special and general).
The picture of a ``time-evolving'' universe seems particularly unappealing when the universe
violates the condition of global hyperbolicity, 
which, indeed, is the case with completely evaporating black holes (see Fig.~\ref{fig1}).
Instead, one of the main messages of the theory of relativity is that time should be treated
on an equal footing with space. In particular, it seems natural to adopt
the {\em block time} (also known under the name block universe; see, e.g., \cite{ellis}
and references therein)
picture of the universe, according to which the universe does {\em not evolve} with time,
but is a ``static'' 4-dimensional object in which ``past'', ``presence'', and ``future'' 
equally exist. For example, such a view automatically resolves causal paradoxes 
associated with closed causal curves \cite{nikcaus}.

The basic intuitive idea how the block-time picture of the universe resolves the black-hole information paradox can be seen from Fig.~\ref{fig1} (see also \cite{hartle}). 
From the standard point of view, only the outgoing particle exists in the far future, while the ingoing particle is destroyed. Consequently, information
encoded in the correlations between outgoing and ingoing particles is destroyed. On the other hand,
from the block-time point of view the past also exists, so the information is not destroyed
because the outgoing particle in the far future is correlated with the ingoing particle in the past.
The aim of this paper is to put this intuitive idea into a more precise framework.
In the next section we briefly review the main ideas of the general formalism of treating time in 
quantum theory on an equal footing with space, while the implications on Hawking evaporation are discussed in Sec.~\ref{SEC3}.

\section{Treating time in QM on an equal footing with space}
\label{SEC2}

The first step towards treating time on an equal footing with space in quantum mechanics (QM)
is to extend the probabilistic interpretation of a 1-particle wave function 
$\psi({\bf x},t)\equiv\psi(x)$ \cite{stuc,niktime}. Instead of the usual infinitesimal probability
of finding particle at the space position ${\bf x}$ 
\begin{equation}\label{p1}
 dP_{(3)}=|\psi({\bf x},t)|^2 d^3x ,
\end{equation}
one has the infinitesimal probability of finding particle at the spacetime position $x$
\begin{equation}\label{p2}
 dP=|\psi(x)|^2 d^4x .
\end{equation}
The usual probability (\ref{p1}) is then recovered from (\ref{p2}) as a special case,
corresponding to the {\em conditional} probability that the particle will be found
at ${\bf x}$ if it is already known that it is detected at time $t$. More precisely, 
since $\psi$ in (\ref{p1}) and (\ref{p2}) do not have the same normalizations, 
the variant of (\ref{p1}) that emerges from (\ref{p2}) should be written as
\begin{equation}\label{p3}
 dP_{(3)}=\frac{|\psi({\bf x},t)|^2 d^3x}{N_t} ,
\end{equation} 
where 
\begin{equation}\label{p4}
N_t=\int |\psi({\bf x},t)|^2 d^3x 
\end{equation}
is the normalization factor. As discussed in \cite{niktime}, such a generalized probabilistic 
interpretation allows to define the time operator in QM, 
solves the problem of probabilistic interpretation of solutions to
the Klein-Gordon equation, 
and provides a better explanation of the standard rule 
that transition amplitudes should be interpreted in terms of transition probabilities
{\em per unit time}.

The next step is to generalize this to the case of many-particle wave functions.
To treat time on an equal footing with space, one needs to introduce a many-fingered time
wave function \cite{tomon}. A state describing $n$ particles
is described by the many-fingered time wave function 
$\psi({\bf x}_1,t_1, \cdots , {\bf x}_n,t_n)\equiv\psi(x_1,\ldots,x_n)$. Consequently, (\ref{p2}) generalizes to \cite{niktime}
\begin{equation}\label{e6n}
dP=|\psi(x_1,\ldots,x_n)|^2 d^4x_1 \cdots d^4x_n .
\end{equation}
In particular, if the first particle is detected at $t_1$, second particle at $t_2$, etc., 
then Eq.~(\ref{p3}) generalizes to
\begin{equation}\label{e7n}
dP_{(3n)}=\frac{|\psi({\bf x}_1,t_1,\ldots,{\bf x}_n,t_n)|^2 d^3x_1 \cdots d^3x_n}
{N_{t_1,\ldots,t_n}} ,
\end{equation}
where 
\begin{equation}\label{e8n}
N_{t_1,\ldots,t_n}=\int |\psi({\bf x}_1,t_1,\ldots,{\bf x}_n,t_n)|^2 d^3x_1 \cdots d^3x_n .
\end{equation}
Indeed, (\ref{e7n}) coincides with the usual probabilistic interpretation of 
the many-fingered time wave function \cite{tomon}. 
The more familiar single-time wave function is a special case corresponding to the 
time-coincidence limit
\begin{equation}
 \psi({\bf x}_1,\ldots,{\bf x}_n;t)=
\psi({\bf x}_1,t_1,\ldots,{\bf x}_n,t_n)|_{t_1=\cdots=t_n\equiv t} \; .
\end{equation}
In this case (\ref{e7n}) reduces to the familiar single-time probabilistic interpretation
\begin{equation}\label{e7ns}
dP_{(3n)}=\frac{|\psi({\bf x}_1,\ldots,{\bf x}_n;t)|^2 d^3x_1 \cdots d^3x_n}
{N_{t}} ,
\end{equation}
where $N_t$ is given by (\ref{e8n}) at $t_1=\cdots=t_n\equiv t$.

A more difficult step is to generalize this to quantum field theory (QFT), where the number of
particles may be uncertain and may change. The appropriate formalism has recently been 
developed in \cite{nikpilot}. Instead of repeating the whole analysis, 
let us briefly review the final results. In general, a QFT state is described by a 
wave function $\Psi(x_1,x_2,\ldots )$ that depends on an infinite number of spacetime
positions $x_A$, $A=1,2,\ldots,\infty$. Introducing the notation
\begin{equation}
 \vec{x}=\{x_1,x_2,\ldots \} ,
\end{equation}
a QFT state $|\Psi\rangle$ can be represented by the wave function
\begin{equation}
 \Psi(\vec{x})=(\vec{x}|\Psi\rangle 
\end{equation}
satisfying the normalization condition
 \begin{equation}
 \int {\cal D}\vec{x}\, |\Psi(\vec{x})|^2 =1, 
 \end{equation}
where
\begin{equation}\label{vol}
 {\cal D}\vec{x}=\prod_{A=1}^{\infty}d^4x_A .
\end{equation}
Each state can be expanded as
\begin{equation}\label{exp}
 \Psi(\vec{x})=\sum_{n=0}^{\infty} \tilde{\Psi}_n(\vec{x}), 
\end{equation}
where $\tilde{\Psi}_n(\vec{x})$ really depends only on $n$ coordinates $x_A$ and represents
an $n$-particle wave function. The tilde on $\tilde{\Psi}_n$ denotes that this wave function is not normalized. For free fields, i.e., when the number of particles does not change,
the expansion (\ref{exp}) can be written in the form
\begin{equation}\label{exp2}
 \Psi(\vec{x})=\sum_{n=0}^{\infty} c_n \Psi_n(\vec{x}), 
\end{equation}
where $\Psi_n(\vec{x})$ are normalized $n$-particle wave functions
\begin{equation}
 \int {\cal D}\vec{x}\, |\Psi_n(\vec{x})|^2 =1, 
 \end{equation}
and $c_n$ are coefficients satisfying the normalization condition
\begin{equation}
 \sum_{n=0}^{\infty} |c_n|^2 =1 .
\end{equation}
In particular, the vacuum (i.e., the state without particles) 
is represented by a constant wave function
\begin{equation}
\Psi_0(\vec{x})=\frac{1}{\sqrt{{\cal V}}} , 
\end{equation}
where ${\cal V}$ is the volume of the configuration space
\begin{equation}
 {\cal V} = \int {\cal D} \vec{x} . 
\end{equation}
The probabilistic interpretation of (\ref{exp}) is given by a natural generalization of
(\ref{e6n})
\begin{equation}\label{e6ng}
{\cal D}P=|\Psi(\vec{x})|^2 {\cal D}\vec{x}.
\end{equation}

By using the techniques developed in \cite{nikpilot}, the wave functions 
$\tilde{\Psi}_n(\vec{x})$ can in principle be calculated for any interacting QFT.
These wave functions contain a complete information about probabilities
of particle creation and destruction. Let us briefly discuss how this probabilistic
interpretation works.
Let $x_{n,1},\ldots, x_{n,n}$ denote $n$
coordinates $x_A$ on which 
$\tilde{\Psi}_n(\vec{x})\equiv \tilde{\Psi}_n(x_{n,1},\ldots, x_{n,n})$
really depends. (With respect to other coordinates $x_A$, $\tilde{\Psi}_n(\vec{x})$ is a 
constant.) 
If $\tilde{\Psi}_n(x_{n,1},\ldots, x_{n,n})$ vanishes for $x^0_{n,a_n}=t$,
then the probability that the system will be found in the
$n$-particle state at time $t$ vanishes. If $\tilde{\Psi}_n(x_{n,1},\ldots, x_{n,n})$
does not vanish for $x^0_{n,a_n}=t'\neq t$, then there is a finite probability
that the system will be found in the
$n$-particle state at time $t'$. This corresponds to a probabilistic description of
particle creation or destruction
when $t'>t$ or $t>t'$, respectively.
As shown in \cite{nikpilot}, for coincidence times
the probabilities obtained this way coincide with those
obtained by more conventional single-time methods in QFT. Thus, the many-fingered
time formalism is not really a modification, but only an extension of the conventional
QFT formalism.

\section{Implications on unitarity of Hawking evaporation}
\label{SEC3}

Now let us discuss how such a general formulation of QFT enriches our understanding
of Hawking evaporation. Unfortunately, the explicit calculation of $\Psi(\vec{x})$
describing the Hawking evaporation is prohibitively difficult. Nevertheless, 
some qualitative features of $\Psi(\vec{x})$ can easily be inferred from the standard
results \cite{hawk1}. It turns out that these qualitative features are sufficient to understand
how the description of Hawking evaporation by $\Psi(\vec{x})$ resolves
the information paradox. 

For simplicity, we assume that
the set of all particles can be divided into ingoing particles
that never escape from the horizon and outgoing particles that go to the future infinity
(Fig.~\ref{fig1} shows a pair of such particles).
Therefore, all these particles are described by a wave function of the form
\begin{equation}\label{wf}
 \Psi(\vec{x})=\Psi(\vec{x}_{\rm in},\vec{x}_{\rm out}) .
\end{equation}
Since the Hawking particles are created in pairs, this wave function can be expanded as
\begin{equation}\label{wf2}
 \Psi(\vec{x}_{\rm in},\vec{x}_{\rm out}) = \sum_{n=0}^{\infty}
\tilde{\Psi}_{2n}(\vec{x}_{\rm in},\vec{x}_{\rm out}) ,
\end{equation}
where $\tilde{\Psi}_{2n}(\vec{x}_{\rm in},\vec{x}_{\rm out})$ really depends 
on $n$ ``ingoing coordinates'' $x_{{\rm in}\,A}$ and 
$n$ ``outgoing coordinates'' $x_{{\rm out}\,A}$.
(In fact, the wave functions $\tilde{\Psi}_{2n}$ depend also on $\vec{x}_{\rm back}$
describing the background particles of initial black-hole matter,
but for the sake of notational simplicity the dependence on 
$\vec{x}_{\rm back}$ is suppressed.) 
The fact that the state is initially in the vacuum means that
all $\tilde{\Psi}_{2n}(\vec{x}_{\rm in},\vec{x}_{\rm out})$ with $n\geq 1$ vanish for small 
values of $x^0_{{\rm in}\,A}$ and $x^0_{{\rm out}\,A}$. 

The wave function (\ref{wf2}) is a pure state. It describes the whole system of ingoing and 
outgoing particles for all possible values of times of each particle. 
The correlations between all these particles can also be described by the density matrix
\begin{equation}\label{matrpure}
\rho(\vec{x}_{\rm in},\vec{x}_{\rm out}| \vec{x}'_{\rm in},\vec{x}'_{\rm out}) =
\Psi(\vec{x}_{\rm in},\vec{x}_{\rm out})
\Psi^*(\vec{x}'_{\rm in},\vec{x}'_{\rm out}) , 
\end{equation}
which is nothing but a density-matrix representation of the pure state  (\ref{wf2}).
However, an outside observer
cannot detect the inside particles. Consequently, his knowledge is described 
by a mixed state obtained by tracing out over unobservable ingoing particles
\begin{equation}\label{mix}
 \rho_{\rm out}(\vec{x}_{\rm out}|\vec{x}'_{\rm out})=
\int {\cal D}\vec{x}_{\rm in} \; 
\rho(\vec{x}_{\rm in},\vec{x}_{\rm out}| \vec{x}_{\rm in},\vec{x}'_{\rm out}) .
\end{equation}
(Of course, since now we work in a curved background, the measure (\ref{vol}) 
is now modified by the replacement $d^4x_A\rightarrow \sqrt{|g(x_A)|}\, d^4x_A$.)
Nevertheless, the whole system is still described by the pure state (\ref{matrpure}).

Now we are ready to discuss how our approach resolves the information paradox.
For convenience, we choose the global time coordinate such that equal-time hypersurfaces
correspond to (undrawn) horizontal lines in Fig.~\ref{fig1}.
Let us assume that the complete evaporation ends at time $T$, after which
neither a black hole nor a remnant is present.
From the standard semi-classical analysis \cite{hawk1}, we know that ingoing 
particles have zero probability of being found at times larger than $T$.
They are destroyed at the singularity that does not exist for times after the complete evaporation,
as illustrated by Fig.~\ref{fig1}. 
Nevertheless, the pure state (\ref{wf2})
is well defined for all values of $x^0_{{\rm out}\,A}>T$. But what happens if we put 
$x^0_{{\rm in}\,A}>T$? For such values of $x^0_{{\rm in}\,A}$
the wave function (\ref{wf2}) is still well defined, 
but the value of $\Psi$ turns out to be equal to zero, because the probability
of finding the ingoing particles at $x^0_{{\rm in}\,A}>T$ is zero.
A wave function with the value zero does not encode much information,
which corresponds to an apparent loss of information at times larger than $T$.
Still, a wave function with zero value is still a wave function, so the state is still pure.
In fact, since only outgoing particles are present for times larger than $T$, there is no
much point in considering the case $x^0_{{\rm in}\,A}>T$.
To obtain a nontrivial information from (\ref{wf2}) at times larger than $T$, one should
only put $x^0_{{\rm out}\,A}>T$, while times of ingoing particles should be
restricted to $x^0_{{\rm in}\,A}<T$. In that case, the pure state (\ref{wf2})
describes how the outgoing particles at times after the complete evaporation are correlated with
the ingoing particles before the complete evaporation. Such nonlocal correlations
cannot be measured by local observers that cannot travel faster than light, so
information seems lost from the point of view of local observers. 
Nevertheless, these correlations are still encoded in the total wave function 
of the universe, so the principles of QM are not violated -- the wave function of the
universe is still pure. 

Thus, we see how treating time on an equal footing with space provides a new, 
purely kinematic solution to the black-hole information paradox,
without need to understand the details of dynamics. Essentially, 
the block-time picture of the universe
makes any time-dependent problem in 3 spacial dimensions analogous to a time-independent
problem in 4 spacial dimensions. Consequently, there can be no fundamental problem with
non-unitary evolution of the quantum state simply because the concept of evolution itself
does not have any fundamental meaning. Instead, all we have are correlations
among particles at different spacetime positions. Thus, 
even if the original Hawking calculation \cite{hawk1} is essentially correct (in the sense 
that the black hole eventually evaporates completely and that the outgoing radiation 
cannot be described by a pure state), the information is still there, encoded in the 
correlations between outgoing particles in the future and ingoing particles in the past.
From this point of view, the original Hawking calculation may be essentially correct, 
but it is not complete because it only describes correlations among particles at the same
spacelike hypersurface.\footnote{The fact that the outgoing and the ingoing particle 
in Fig.~\ref{fig1} can be connected by a 
spacelike hypersurface does not help. The outgoing particle can interact with other
outside particles, which may transfer information to outside particles that cannot be connected
with the ingoing particle by a spacelike hypersurface.}

To conclude, we believe that our results represent a new step towards
reconciliation of quantum mechanics (QM) with general relativity (GR).
GR suggests that time should be treated on an equal footing with space,
while QM demands unitarity. We have shown that the former (i.e., treating time on an equal footing
with space) automatically restores the latter (the unitarity).

\section*{Acknowledgements}

This work was supported by the Ministry of Science of the
Republic of Croatia under Contract No.~098-0982930-2864.

\end{document}